1

# Filamentary superconductivity of resistively-switched strontium titanate

K. Szot[1,2], C. Rodenbücher[3,*], K. Rogacki[4], G. Bihlmayer[5,6], W. Speier[6], K. Roleder[1], F. Krok[7], H. Keller[8], A. Simon[9] and A. Bussmann-Holder[9]

[1] A. Chełkowski Institute of Physics, University of Silesia, 41-500 Chorzów, Poland
[2] aixACCT Systems GmbH, 52068 Aachen, Germany
[3] Institute of Energy and Climate Research (IEK-14), Forschungszentrum Jülich GmbH, 52425 Jülich, Germany
[4] Institute of Low Temperature and Structure Research, PAS, 50-050 Wrocław, Poland
[5] Peter Grünberg Institute (PGI-1), Forschungszentrum Jülich GmbH, 52425 Jülich, Germany
[6] JARA-FIT and Peter Grünberg Institute (PGI-SO), Forschungszentrum Jülich, 52425 Jülich, Germany
[7] M. Smoluchowski Institute of Physics, Jagiellonian University, 30-348 Kraków, Poland
[8] Physik-Institut der Universität Zürich, 8057 Zürich, Switzerland
[9] Max-Planck-Institute for Solid State Research, 70569 Stuttgart, Germany

Correspondence to c.rodenbuecher@fz-juelich.de

**Abstract**

$SrTiO_3$, although a wide gap insulator, has long been known to become metallic and superconducting at extremely low doping levels. This has given rise to questions concerning the coexistence or interdependence of metallicity, superconductivity, and the material's polar properties. This issue becomes especially intriguing in conjunction with the observation that filamentary metallicity can be induced by means of resistive switching at conditions well below relevant doping levels for homogeneous metallicity. In this study, we demonstrate that resistive switching can also be employed to generate superconductivity at the superconducting transition temperature of $T_c \approx 0.2$ K in $SrTiO_3$. By combining local characterization of the conductivity with theoretical analysis, we propose that the superconducting properties are associated with the electro-formation of columnar-like bundles with a diameter of 40–50 nm, consisting of metallic filaments surrounded by polar regions. We provide a theoretical model identifying the coexistence of metallic and polar regions as a prerequisite for the filamentary-like superconductivity observed.

**Introduction**

The electrical manipulation of nanoscale transport phenomena has enabled the development of resistive switching devices with enormous application potential for microelectronics, e.g., neuromorphic computing[1]. The first important step in the fabrication of a resistively-switchable structure is the formation of filaments by means of electro-reduction/-degradation[2–4]. This local introduction of inhomogeneity by oxygen vacancies enables to engineer the properties, from insulating, to semiconducting or metallic, by controlling the size and distribution of the nanofilaments. For an $SrTiO_3$ (STO) single crystal, a model material for transition metal perovskite oxides, the evolution of filaments in  conjunction with electroforming does not occur randomly in the matrix, but is predetermined by the existence of a dislocation network in the surface region[5]. The extraordinary crystallographic geometry[6], electronic defect structure[7], flexoelectric or ferroelectric polarization[8], magnetic properties[9], and local chemical composition[10] of the dislocation cores combined with the invariance of the Burgers vector[11], induce the channeling of the current flow along the dislocation network within the non-conducting/dielectric matrix of STO. Experimentally, it has been proven by local-conductivity atomic force microscopy (LC-AFM) that the electrochemical or thermal deoxidation of STO crystals is restricted to dislocation cores[3,12,13]. This observation is consistent with the theoretical calculation of the formation enthalpy of oxygen vacancies close to dislocations, which is significantly lower than in the matrix[14]. As a result of the preferential removal of oxygen, the regions close to the dislocation cores are enriched by oxygen vacancies leading to a local phase transformation towards $TiO_x$ suboxides providing excess $d^1$ electrons, which are necessary for metallic conductivity[15]. In this manner, the idea of generating conducting nanowires along dislocations in insulating materials[16] can be easily realized; in our case, however, by means of electrically-induced self-doping. Through thermographic studies of crystalline samples with artificially-introduced bands of extended defects connecting the cathode with the anode, the preferential current channeling along the predefined dislocation network during electro-reduction has been experimentally-illustrated[5]. It has been shown that transformation into the metallic state occurs at temperatures of 300–500 °C, which is too low to permit the thermal removal of oxygen[3]. Nevertheless, an effusion of oxygen during electro-reduction has been detected via mass spectrometry[15,17]. Accordingly, such an electro-deoxidation effect changes the chemical composition of the oxide locally, as has already been observed in various metal oxides[18–21]. During electro-formation, oxygen ions move within the electric field from the cathode towards the anode[22]. Hence, the evolution of metallic filaments starts at the cathode, shifting a virtual cathode through the crystal until a galvanic shortcut is achieved. It must be noted that such a short-circuit should not be mistakenly associated with a macroscopic transition of the "whole crystal" into a metal[23].

Superconductivity in STO has been well known since many years, and has been induced by (nominally) homogeneous doping using extrinsic donors or self-doping by oxygen vacancies upon thermal reduction[24–26], by modulating the charge carrier density via transistor-like configurations[27], by generating two-dimensional electron gases, e.g., in $LaAlO_3/SrTiO_3$ structures or on modified STO surfaces[28,29], as well as by exploiting interface effects, e.g., in FeSe/STO structures[30]. Early on, the superconductivity in STO has been associated with the long wave length transverse soft optic (the ferroelectric) mode. With the discovery of superconductivity at even lower doping levels, this viewpoint has again been adopted, in addition to novel approaches that have been suggested that are beyond BCS theory. Here, we demonstrate that electro-formation can be employed to generate metallic filaments, which are clustered in bundles with diameters of 40–50 nm. The regions between the filaments, the embedding matrix, are polar in character. Electro-reduced STO is therefore a highly inhomogeneous material with coexisting charge-rich and charge-poor regions, i.e., metallicity with insulating properties being intimately entangled.

## Results

### Electro-reduction

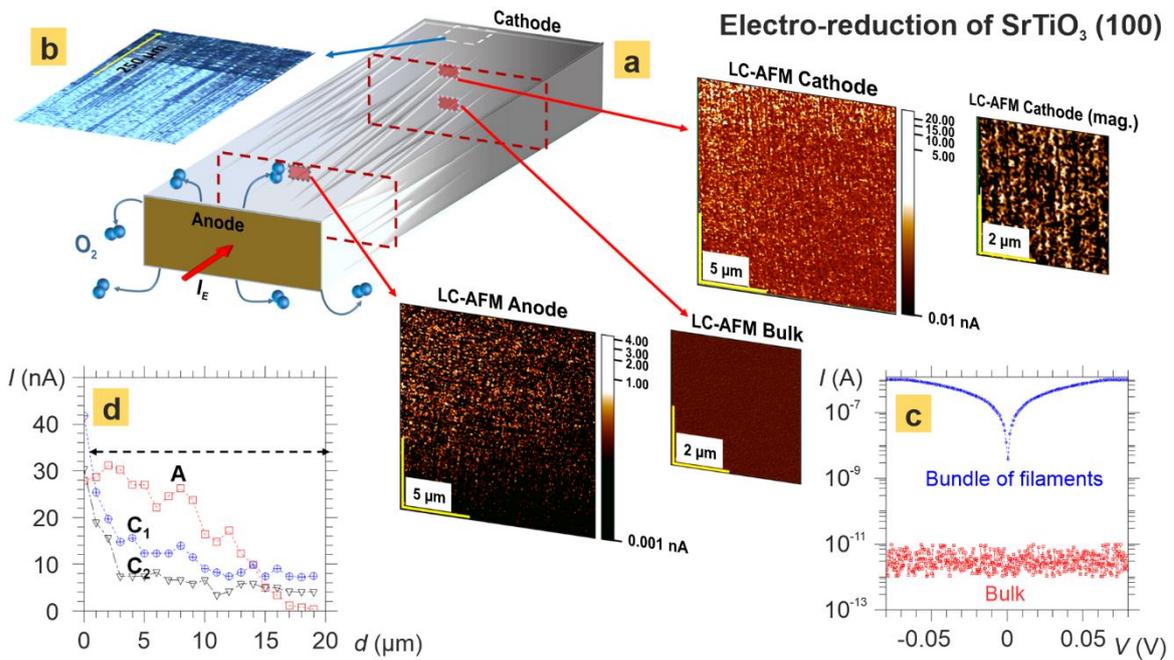

**Figure 1 Microscopic view on the electro-reduction of STO. a**) Schematic view of the experimental situation for electro-reduction of STO with cross-sectional LC-AFM maps obtained after cleaving the crystal in the anode and cathode regions (with magnification, mag.); **b**) micrograph of the distribution of dislocations in the cathode region revealed by the etch-pits technique; **c**) $I/V$ measurements obtained by LC-AFM on a bundle of filaments and in the bulk region; **d**) average current as a function of the distance to the surface as calculated from the LC-AFM maps close to the anode (A) and in two positions close to the cathode (C1, C2).

An STO (100) single crystal was subjected to electro-reduction under vacuum conditions ($I = 10$ mA, $t = 2$ h, $p = 10^{-7}$–$10^{-8}$ mbar, $T = 350$ °C; for details, see the supplementary information). After the macroscopic resistance of the sample had dropped by many orders of magnitude, indicating metallic properties, the crystal was investigated by means of LC-AFM and optical microscopy, as shown in Figure 1. The LC-AFM maps obtained in the cathode region after cleaving the crystal (Figure 1a) and the optical micrograph of the exits of dislocations as revealed by etch-pits technique (Figure 1b) reveal the inhomogeneous distribution of filamentary structures following electro-reduction. Close to the cathode, a higher concentration of electro-reduced filaments was present than in the vicinity of the anode due to the electro-migration of the oxygen ions in the electrical field gradient. The LC-AFM maps exhibit a hierarchical character of the conducting filaments in the surface region, similar to the filaments in the surface region of thermally-reduced crystals, thus supporting the idea that dislocations predefine the shape of the filamentary network. In all regions of the sample, well-conducting filaments only existed in the last few tens of micrometers of the surface region, whereas the cross-sectional LC-AFM maps of the bulk do not show conducting filaments, and so evidence that the bulk of the crystal remains insulating. The lack of electrical conductivity can also be identified for the matrix region between the filaments if the distance between the filaments is larger than a few nanometers. This is indicated by I/V measurements obtained by contacting either a filament or matrix with the LC-AFM tip (Figure 1c). The results for the filaments exhibit a linear current–voltage dependence (which is curved in the logarithmic representation), whereas the current in the bulk region was below the detection limit. To illustrate the depth dependence of the conductivity, the mean current flow at the cathode (C1 and C2) and anode (A) is presented in Figure 1d. Each point of the curves was calculated by integrating the current of an "LC-AFM strip" with an area of $5 \times 1$ µm$^2$, starting from the surface to the strip at a distance of 19 µm below the surface. Close to the surface, the averaged current was highest, and then dropped significantly within a few micrometers, revealing the "skin-like character" of the electric transport, with the electro-formed surface region carrying the lion's share of the total macroscopic current. Using a grain analysis, the dimension of the conducting spots in the LC-AFM maps was determined to be approximately 45 nm.

A strong agglomeration of filaments into orthogonal bands in the [100] direction was present near the cathode as can be seen as stripe-like structures, both in the etch-pits and the LC-AFM maps of Figure 1. The LC-AFM maps indicate that the filaments do not occur as single filaments isolated from each other, but form filament bundles. In fact, the LC-AFM maps indicate that the filaments do not occur as single filaments isolated from each other but form filament bundles. This bundling effect was also observed in thermally-reduced crystals[13], but is much more pronounced for electro-reduced crystals. In order to investigate the arrangement of the filaments in more detail, the conducting spots were measured in a higher magnification. Therefore, a region with a high density of dislocations close to the cathode was identified using the etch-pits technique, as is shown in Figure 2a. An arrangement of the etch-pits marking the exits of the dislocations in the [100] direction, perpendicular to the cathode, was observed, which confirmed the agglomeration of filaments in the form of a linear band. In order to investigate the distribution of the filaments within the band, LC-AFM maps were made for the surface close to the cathode (Figure 2b). The current map shows that the filaments were not statistically distributed within the bands, but were instead clustered in bundles. From the magnified LC-AFM maps (Figure 2c, 2d) it can be seen that the bundles with a radius of 40–45 nm had a discrete conductivity character and consisted of nanofilaments with radii of approximately 2 nm (Figure 2e). Using the simple statistics obtained for a few bundles (Figure 2c), the average number of nanofilaments per bundle was estimated to be 200–300. Figure 2d shows that between the highly conducting spots marking the nanofilaments, an increased conductivity can also be observed, which is one-to-two orders of magnitude lower than in the center of the filaments, but still significantly higher than in the insulating matrix. For most of the nanofilaments in a bundle, the distance to the neighboring nanofilament was less than 3–4 nm, suggesting that the nanofilaments act as collective electron sources that dope the region between themselves. In contrast to this, the conductivity between a nanofilament and matrix at the edge of a bundle is three orders of magnitude smaller, within only 1 nm (Figure 2e). This is in agreement with the DFT calculations of $TiO_2$ that predict a highly confined spatial spread of the electron density near an oxygen vacancy-rich filament[31]. In order to gain more insight into the structure of the nanofilaments themselves, one nanofilament was mapped using LC-AFM with atomic resolution (Figure 2f), which displayed a direct correlation to the dislocations. Close to the exit of a dislocation, whose position could be identified by analyzing the course of the atomic rows, the local conductivity reached a maximum. This behavior supports the hypothesis that the dislocation core is slightly reduced, or in our case electro-reduced by the local removal of oxygen, which is equivalent to strong local doping. Moreover, the temperature dependence of the conductivity of the selected sets of bundles measured by LC-AFM (Figure 2g) exhibits an increasing resistance and thus metallic behavior, which is typical, e.g., for low $TiO_x$ phases. In summary, the high-resolution LC-AFM investigation reveals that the conducting spots that were seen in Figure 1 and 2 are, in fact, bundles of highly conductive nanofilaments with a self-doped region between them, and can be considered "thick nanoscopic wires".

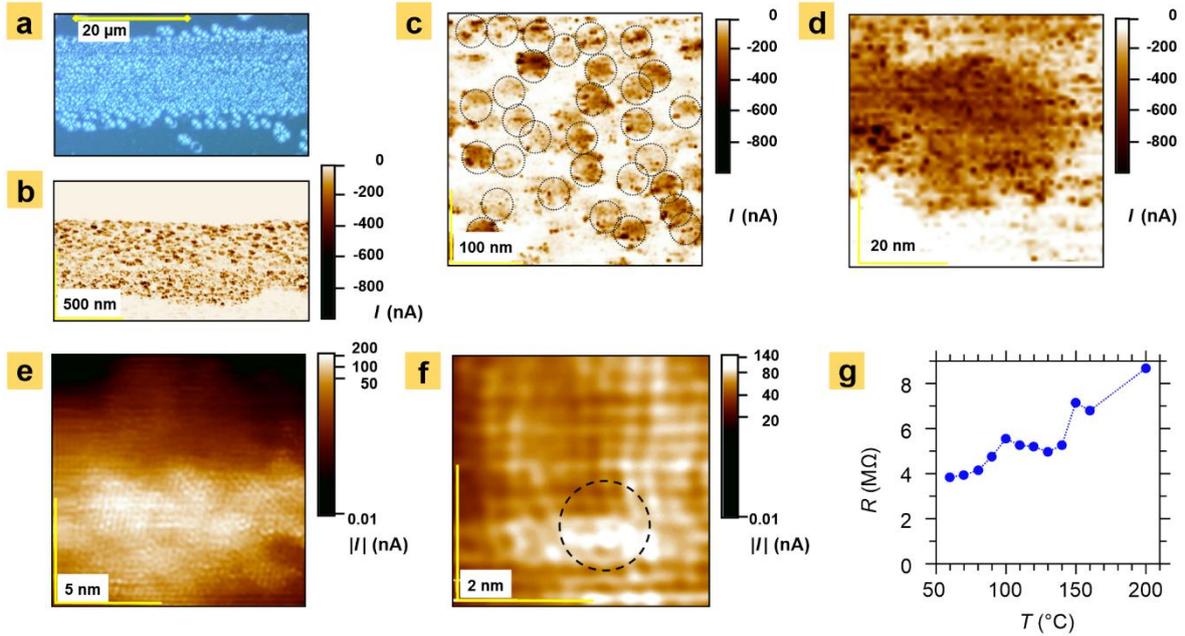

**Figure 2 Detailed analysis of the dimension of the filaments. a**) Microscopic inspection of the etch pit distribution marking dislocations in a band along the <100> direction; **b**) in-plane LC-AFM map obtained near the cathode ($U_{sample}$ = -0.01 V); **c**) magnification of b) with filament bundles of 40-50 nm marked by a black dashed line; **d**) LC-AFM map showing the discrete structure of conductivity in a typical bundle. The very good conducting filaments and weaker conducting regions of the matrix are visible between the closely located filaments in the bundle; **e–f**) LC-AFM maps of a nanofilament obtained with atomic resolution; and **g**) temperature dependence of the resistance, measured by placing the LC-AFM tip above a nanofilament.

**Simulation of the electronic structure**

In order to shed some light on the nature of metallic filaments forming in an insulating perovskite matrix, we performed DFT calculations (for computational details see Al-Zubi et al.[32]) of extended defects, as is shown in Figure a. A one-dimensional, stoichiometric defect creates an "inner surface" where an oxygen vacancy ($V_O$) leads to the formation of a defect state at the bottom of the conduction band. As the formation energy of $V_O$ is lower at an (inner) surface than in the bulk[33], the accumulation of such defects at inner surfaces is likely. In addition, tendencies for the clustering of these defects have been reported[34]. The charge density of the defect states is primarily located on the nearby Ti ion (the blue DOS curve in Figure a) and in the region between the atomic spheres (red line). In Figure b, we estimated the charge density locally by integrating the induced defect charges in the Ti atomic spheres of the individual unit cells (red bars). A charge density of the order of $5 \times 10^{20}$ electrons per cm$^3$ extends into the region between two such extended defects (columns 1 and 6). In the matrix, this defect charge density decays quickly, but in a bundle of one-dimensional defects, larger conductive areas can form.

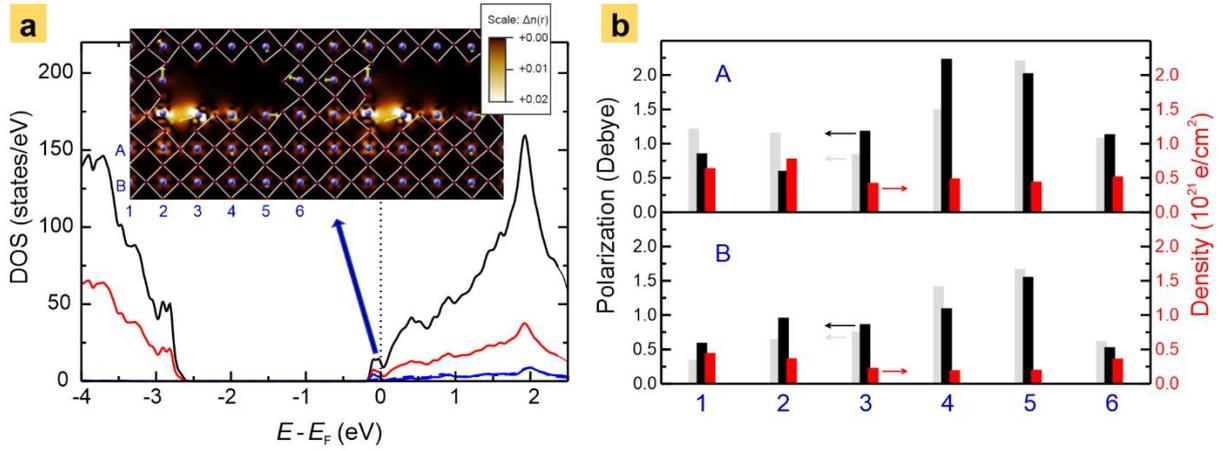

**Figure 3 DFT simulation of neighboring one-dimensional extended defects in STO. a**) Illustration of the electronic structure. The relaxed structure is shown in the inset with oxygen atoms marked in red and Ti atoms in blue. The vacancy-induced states are visible in the density of states (DOS) at the bottom of the conduction band. The total DOS is shown in black, the local DOS on the Ti atoms adjacent to the vacancy in blue and the interstitial DOS (between the atomic spheres) in red. A cut through the induced charge density is shown in the inset underlying the atomic structure. Bright areas indicate a high electronic density (see thermometer inset). Large local polarizations are indicated by yellow arrows. **b**) Polarization (black) and defect-induced charge densities (red) in STO near the extended defect. The vertical and horizontal positions of the Ti atoms are labeled with (A,B) and (1,…,6), respectively, and are visualized in the inset of (a). For comparison, the size of the local polarizations in an extended defect without the oxygen vacancy are indicated by the gray bars.

We investigated how these local charges influence the local, static polarizations as the formation of polar regions will be essential for superconductivity. To estimate this quantity, we employed a simple ionic model based on the formal charges and the positions of O ($r_O$) and Sr ($r_{Sr}$) relative to Ti ions, calculating $P_{Ti}=\sum(r_{Sr}/4-r_O)$. These polarizations are indicated by yellow arrows in the inset of Figure a and their norm is summarized in Figure b (black bars). We compare them to the polarization of a stoichiometric extended defect without $V_O$ (grey bars). As is apparent, a small electronic density (below $5 \times 10^{20}$) does not quench the local polarizations (of course, some reconstruction also results from the presence of the additional defect), and larger densities (position B1 and B2) can lead to some local reduction of $P_{Ti}$.

**Measurements of the superconducting properties**

Having induced a network of bundled-conducting nanofilaments in the surface region of STO by electro-reduction, we analyzed the superconducting properties via low-temperature conductivity measurements in a magnetic field. For comparison, we also investigated a crystal, in which a network of nanofilaments was generated by means of pure thermal annealing and thus was not as pronounced and bundled as for the electro-reduced one[13]. As our previous study demonstrated that the progression of deoxidation could differ in crystals from different manufacturers, we utilized samples from the same crystal to ensure comparability. For the thermally- and electro-reduced STO samples described above, the temperature dependencies of the resistance were measured, and the results are shown in Figure 4a. The resistance was divided by the distance between voltage contacts, $R/l$, such that for samples with the same cross-section, the linear resistance characterizing a given material was obtained. As is evident from Figure 4a, $R/l$ decreases with temperature designating metallic properties. The resistance of the electro-reduced sample was significantly lower than that of the thermally-reduced one, indicating that the increasing density of metallic nanofilaments in the electro-reduced sample correlates with the macroscopic resistance of the sample. This confirms that the metallic properties of deoxidized STO are, as we demonstrated, determined by the nanofilaments.

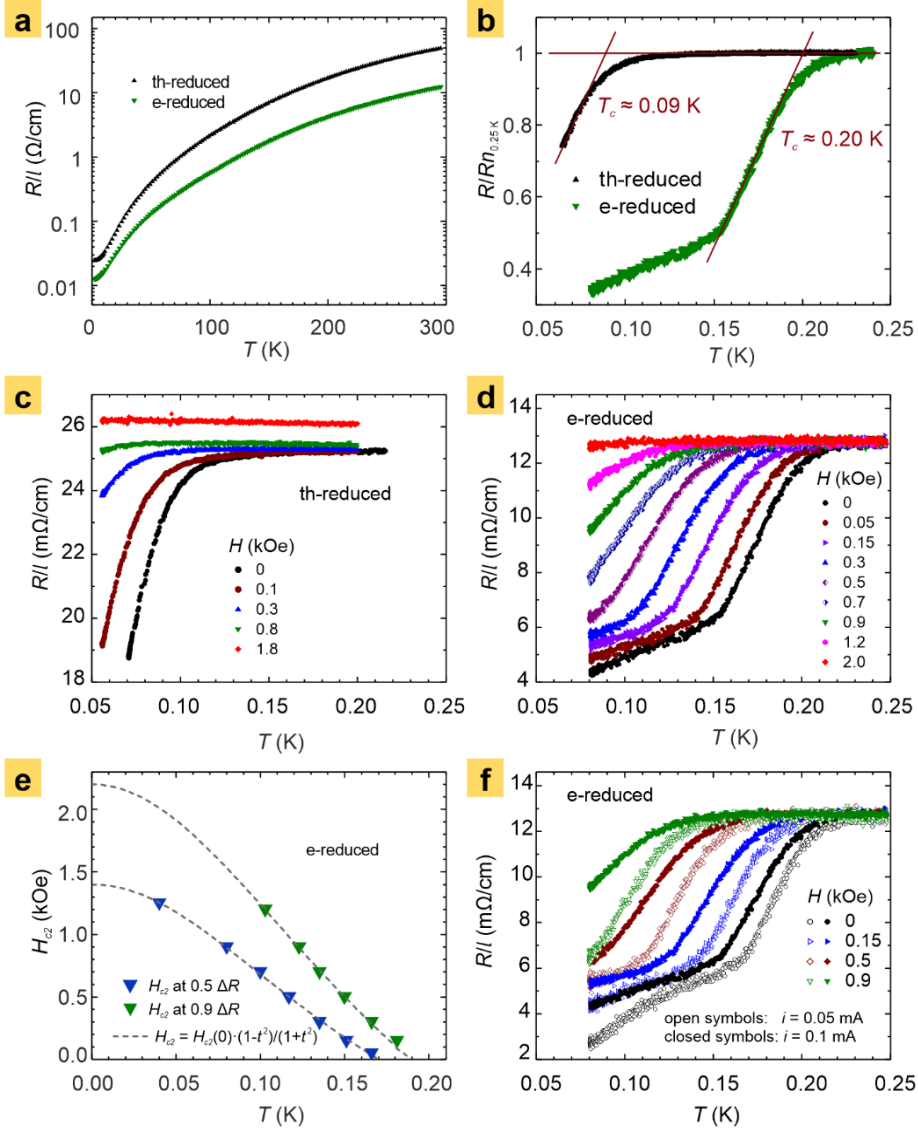

**Figure 4 Resistance measurements of the thermally- and electro-reduced STO samples. a**) Resistance, $R/l$, vs. temperature for the thermally- (th-reduced) and electro-reduced (e-reduced) STO samples with the same cross-sectional area, where $l$ is the distance between the voltage leads; **b**) $R/Rn_{0.25K}$ ratio for the thermally- and electro-reduced STO samples below the superconducting transition temperature $T_c$ ($Rn_{0.25K}$ is the value of $R$ in the normal state at 0.25 K). Note that in the electro-reduced STO sample, $T_c$ is considerably larger than that for the thermally-reduced one; **c**) resistance, $R/l$, vs. temperature for the thermally-reduced STO sample measured in different magnetic fields at a current of $I = 0.1$ mA; **d**) resistance, $R/l$, vs. temperature for the electro-reduced STO sample measured in different magnetic fields at a current of $I = 0.1$ mA; **e**) upper critical field, $H_{c2}$, vs. temperature for the electro-reduced STO sample. The results were obtained using the data displayed in d) with the criterion $0.5\Delta R$ (open triangles) and $0.9\Delta R$ (closed triangles), where $\Delta R$ is the change in the resistance at the transition to the superconducting state – the dashed lines show the fitting of the experimental results to the equation $H_{c2}(T) = H_{c2}(0)(1-t^2)/(1+t^2)$ where $t=T/T_c$; **f**) resistance, $R/l$, vs. temperature for the electro-reduced STO sample measured in different magnetic fields and at two measuring currents: $I = 0.05$ mA (open symbols) and $I = 0.1$ mA (closed symbols) for comparison.

In the temperature regime below 0.25 K, both samples exhibit a sudden drop in the resistance, which is characteristic of superconductivity. This is shown in Figure 4b, where the resistance $R$ normalized to the normal-state resistance $Rn_{0.25K}$ at 0.25 K is presented in the temperature region of the transition temperature, $T_c$. The superconducting transition temperature for the electro-reduced sample ($T_c \approx 0.2$ K), is more than a factor of 2 larger than that for the thermally-reduced sample ($T_c \approx 0.09$ K), suggesting that superconductivity in the electro-reduced sample is caused by the bundled nanofilaments, and can be enhanced by their increased density. An increase in $T_c$ with the volume of the superconducting phase has been observed for "nano-sized" materials, e.g., in In and Pb nanoparticles[35] and Mo-Ge nanowires[36], with sizes comparable to the diameters of the nanofilaments (2–5 nm) and filamentary bundles (40–50 nm) found in our samples. Moreover, the resistance of the electro-reduced sample reveals a pronounced hump around 0.15 K (Figure 4b). A similar behavior of the resistance was reported for the ideal inhomogeneous nanofilamentary superconductor, $Na_{2-\delta}Mo_6Se_6$[37]. Note that in the superconducting state, none of the samples exhibited zero resistance, which serves as another strong argument for non-percolative filamentary superconductivity (spatially-separated, phase-coherent superconducting regions) in deoxidized STO[37].

The behavior of $R/l(T)$ in a magnetic field for the thermally- and electro-reduced STO sample is shown in Figure 4c and d, respectively. The results prove the superconducting nature of the transition to a lower resistance state and allow the temperature dependences of the upper critical field, $H_{c2}(T)$ to be determined. Criteria $0.5\Delta R$ and $0.9\Delta R$ were used, where $\Delta R$ is the change in resistance upon entering the superconducting state. Figure 4e shows the $H_{c2}(T)$ results for the electro-reduced sample, with fitting of the experimental data to equation $H_{c2}(T) = H_{c2}(0)(1-t^2)/(1+t^2)$, where $H_{c2}(0)$ is the value of the critical field at $T = 0$, $t = T/T_c$, and $T_c$ is the superconducting transition temperature at $H = 0$. Depending on the criterion, $0.5\Delta R$ or $0.9\Delta R$, the coherence lengths $\xi_0 \equiv \xi(T=0) = [\phi_0/2\pi H_{c2}(0)]^{0.5} \approx 49$ nm and 39 nm were obtained from the values $H_{c2}(0) = 1.4$ kOe and 2.2 kOe, respectively, extracted from the fitted $H_{c2}(T)$ dependences. These $\xi_0$ values are similar to those extracted from the linear extrapolation of $H_{c2}(T)$ to $H_{c2}(0)$ from the last experimental point in the high fields, namely 47 nm and 35 nm ($H_{c2}(0) \approx 1.50$ and 2.73 kOe, for the $0.5\Delta R$ and $0.9\Delta R$ criteria, respectively), which are lower limits for $\xi_0$.

In order to test which of the criteria ($0.5\Delta R$ or $0.9\Delta R$) provides more correct $H_{c2}(T)$ values, the magnetic field dependence of $R(T)$ for the electro-reduced STO sample was analyzed using two different measurement currents (0.05 mA and 0.1 mA). The results are shown in Figure 4f. For both currents, the transition to the superconducting state occurred at the same temperature, but for $I = 0.1$ mA, the transition was significantly broader than for $I = 0.05$ mA, indicating an influence from the vortex dynamics on the transition. In order to minimize this effect, the more appropriate $0.9\Delta R$ criterion was used to extract $H_{c2}$ and $\xi_0$ from the resistance data for the electro-reduced STO sample, yielding $H_{c2}(0) \approx 2.2$ kOe, $\xi_0 \approx 39$ nm, and $T_c \approx 0.20$ K. A coherence length of $\xi_0 \approx 40$–50 nm is typical for conventional low-temperature *bulk* superconductors, such as e.g. Hg, Nb or In. Moreover, the value of $\xi_0 \approx 40$ nm approximately corresponds to the diameter of the filament bundles, 40–50 nm, as obtained from the LC-AFM studies (see Figure 2), supporting our assertion that superconductivity, as well as the metallicity of STO, is filamentary in nature and restricted to the deoxidized cores of dislocations.

**Theoretical description of superconductivity**

Before discussing our model of the superconductivity of the electroformed filament bundles, we consider the conditions that are necessary to induce superconductivity in (self-)doped STO in general. Theoretically, the polarizability model[38–40] and DFT calculations have been employed to confirm the heterogeneous character of doped STO. The polarizability model is based on the nonlinear polarizability of the oxygen ion, i.e., $O^{2-}$, which is unstable as a free ion. In a crystal, the $2p^6$ state is stabilized by the surrounding ions through dynamic covalency, i.e., by sharing an electron with its surrounding neighbors. This implies that the introduction of additional electrons, as in doped or electro-degraded STO, only affects the Ti ions and causes their $d^1$ configuration. In view of the additional charge on Ti, the surrounding $O^{2-}$ ions move away from it and cause locally strongly distorted regions. The essential task for both approaches is, however, to demonstrate the coexistence of a polar matrix with regions that have a different chemical composition and crystallographic structure. By concentrating on the dynamic properties of STO and knowing their development with carrier concentration $n$, the STO-related double-well potential has been self-consistently extracted and has been shown to change its character from double- to single-well with $n$, where $n_c$ defines the border line for this change[23]. As for $n > n_c$, the superconductivity vanishes and a truly homogeneous metallic state is realized, we only concentrate on the region $n < n_c$.

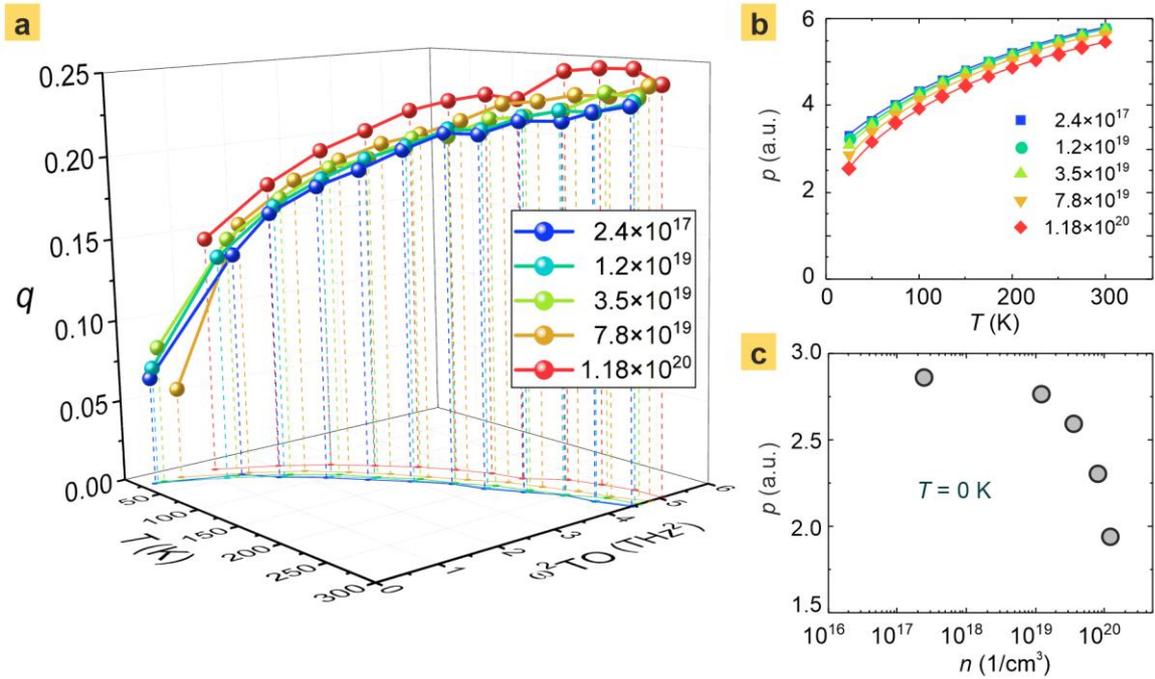

**Figure 5 Lattice dynamics analysis of STO. a**) Temperature and momentum $q$ dependence of the squared soft optic mode frequency $\omega_{TO}^2$ of the STO for different carrier concentrations as indicated by the color code in the figure; **b**) temperature dependence of the local dipole moment $p$ for different carrier concentrations; and **c**) dependence of the dipole moment $p$ on the carrier concentration $n$ for $T = 0$ K.

The essential phonons in the STO are the lowest lying soft transverse optic and acoustic modes. Anomalies in these two branches have been shown to be signatures of the formation of *local polar nano-regions*[41]. By calculating the phonon group velocities for the two considered branches[42], these become very apparent and yield the momentum at which the *local*, spatially-confined soft modes occur, namely the point where the scattering between the two modes is the strongest[43]. The corresponding squared polar optic mode frequency $\omega_{TO}^2(q)$ is displayed in Figure 5a as a function of carrier concentration $n$, temperature $T$ and momentum $q$.

As is obvious from the figure, this *local* polar mode softens with decreasing temperature and simultaneously moves to lower momentum values but never reaches the homogeneous $q = 0$ limit. In addition, the softening is reduced with increasing carrier concentration and its momentum space spread decreases indicating the growing spatial confinement and thus shrinking of these polar nano-domains. In contrast to a long wavelength "true" soft mode, it is nonlinearly dependent on temperature below ≈150 K. As $\omega_{TO}^2(q) \approx \frac{1}{\varepsilon_0}$, the dielectric permittivity $\varepsilon_0$ is extracted from it, and is approximately 40 % smaller than the long wavelength limit, but still exhibits an appreciable temperature dependence, which is typical for an almost ferroelectric compound.

From the calculated thermal average of the displacement-displacement correlation function, a local dipole moment has been derived that shrinks with decreasing temperature and increasing carrier concentration (Figure 5b). The zero temperature limit as a function of the carrier concentration is shown in Figure 5c. As expected, it rapidly decreases with increasing carrier concentration and approaches a constant value for small densities, thus clearly supporting the polar character of the matrix. The carrier concentration range is compatible with our estimates from the DFT calculations shown above.

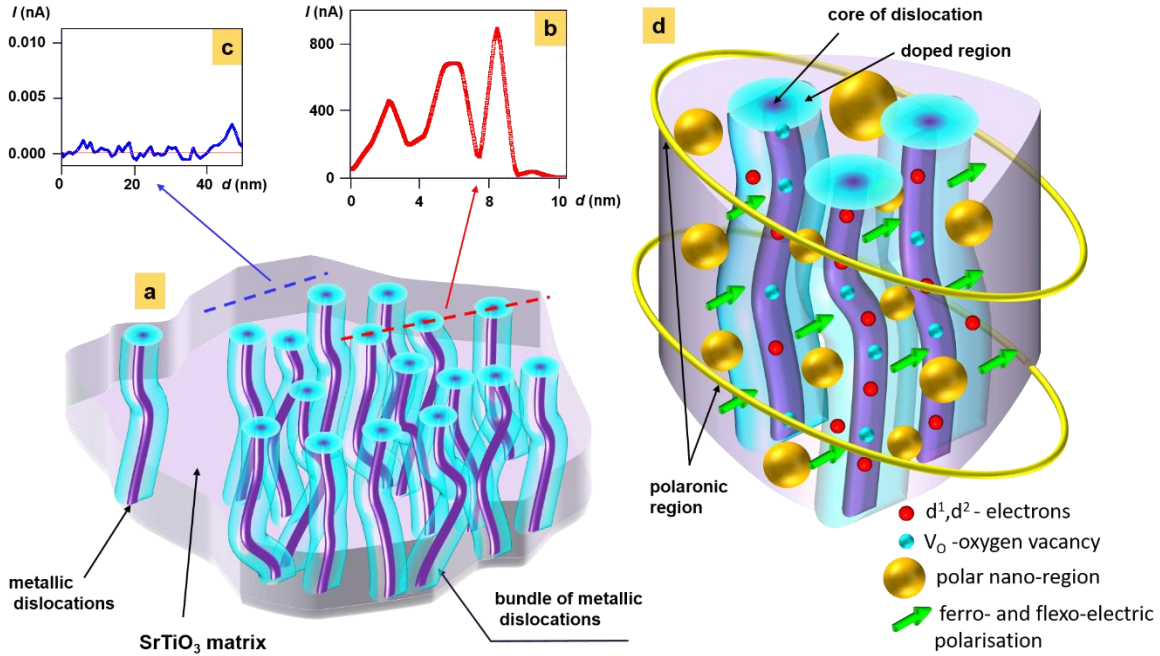

**Figure 6 Schematic picture of the dislocation-based filamentary superconductivity in electro-reduced STO. a**) Arrangement of bundled metallic dislocations according to LC-AFM and etch-pit studies; **b**) LC-AFM line profile across the dislocation cores, revealing that they have an extremely high conductivity, whereas the region between the neighboring dislocation cores exhibits lower but still measurable conductivity; **c**) LC-AFM line profile in the region without electro-reduced dislocations revealing insulating behavior with a current

in the range of the noise of the current-voltage converter; and **d**) illustration of the proposed coupling mechanism of pairs of $d^1$ (and, possibly, $d^2$) electrons in metallic dislocations mediated by a polar nano-region between.

The above analysis demonstrates that in the metallic and superconducting low carrier density limit of STO, metallicity is not globally present but only appears in filaments that coexist with elastically- and polar-distorted domains where the latter shrink in size with increasing carrier density. These have fully vanished beyond a critical carrier density to provide space for a homogeneous metal where superconductivity is absent. The electronic band structure can be interpreted in terms of localized polaronic bands attributed to the matrix, whereas Fermi liquid-type behavior must be present in the filaments. With increasing carrier density, the localized band adopts dispersion from the itinerant one, and is analogous to a steep band/flat band scenario[44] in which superconductivity is a consequence of interband interactions.

**Discussion**

We have shown that the electro-reduction of STO at moderate temperatures ($T < 400$ °C) induces superconductivity with a $T_c \approx 0.20$ K, which is significantly higher than that of thermally-reduced STO ($T_c \approx 0.09$ K). The global removal of oxygen is extremely low and only limited to the cores of the dislocations, where oxygen vacancies are preferentially generated. In this way, bundled metallic filaments form along of the dislocation network, which short-circuit the dielectric matrix and induce macroscopically detectable metallic behavior. Thus, this finding explains why the very low global carrier densities observed in STO can give rise to conductivity and even superconductivity.

The metallic dislocation cores exhibit similar properties to TiO nanowires, and are undoubtedly the carrier of the metallic conductivity above $T_c$, but cannot alone be responsible for the superconductivity. The radius of the filaments (about 2 nm) is smaller than the necessary coherence length found in TiO thin films (about 20 nm)[45] and smaller than the coherence length determined for our electro-reduced crystal by the upper critical field and LC-AFM measurements (40–50 nm).

Therefore, we assume that superconductivity is related to a collective phenomenon involving the metallic cores of dislocations and the polar nano-regions between them. As schematically illustrated in Figure 6, we assume that the d-electrons in the metallic dislocations can be coupled by the polar properties of the STO matrix. This behavior is consistent with the results of the DFT simulations, which indicate that a static polarization in the STO matrix beside a metallic filament is present, although the electrons provided by the filament lightly dope the surrounding matrix. The possibility of generating a polarization in nanoregions at such a doping level was confirmed by lattice dynamic calculations. In other words, the proximity of the metallic filaments does not eliminate the polar character of the STO matrix.

We propose that the interaction of the polar regions of the STO with the $d^1$ and $d^2$ electronic states in different filaments of a bundle constitutes the pairing glue of the Cooper electrons. The determined coherence length perfectly correlates with the average radius of the bundles formed by electro-degradation. As the bundling tendency of the nanofilaments in the electro-degraded crystals is higher than in the thermally-reduced ones, their superconducting properties are also more pronounced, as evidenced by their higher $T_c$.

## Methods

A detailed description of the applied methods is provided in the supplementary information.

## References

1. Yang, J. J., Strukov, D. B. & Stewart, D. R. Memristive devices for computing. *Nat. Nanotechnol.* **8**, 13–24 (2013).

2. Waser, R., Dittmann, R., Staikov, G. & Szot, K. Redox-Based Resistive Switching Memories - Nanoionic Mechanisms, Prospects, and Challenges. *Adv. Mater.* **21**, 2632–2663 (2009).

3. Szot, K., Speier, W., Bihlmayer, G. & Waser, R. Switching the electrical resistance of individual dislocations in single-crystalline $SrTiO_3$. *Nat. Mater.* **5**, 312–320 (2006).

4. Du, H. *et al.* Nanosized Conducting Filaments Formed by Atomic-Scale Defects in Redox-Based Resistive Switching Memories. *Chem. Mater.* **29**, 3164–3173 (2017).

5. Rodenbücher, C. *et al.* Current channeling along extended defects during electroreduction of $SrTiO_3$. *Sci. Rep.* **9**, 2502 (2019).

6. Jia, C. L., Thust, A. & Urban, K. Atomic-scale analysis of the oxygen configuration at a $SrTiO_3$ dislocation core. *Phys. Rev. Lett.* **95**, 225506 (2005).

7. Buban, J. P. *et al.* Structural variability of edge dislocations in a SrTiO$_3$ low-angle [001] tilt grain boundary. *J. Mater. Res.* **24**, 2191–2199 (2009).

8. Gao, P. *et al.* Atomic-Scale Measurement of Flexoelectric Polarization at SrTiO$_3$ Dislocations. *Phys. Rev. Lett.* **120**, 267601 (2018).

9. Shimada, T. *et al.* Ferroic dislocations in paraelectric SrTiO$_3$. *Phys. Rev. B* **103**, L060101 (2021).

10. Du, H., Jia, C.-L. & Mayer, J. Local crystallographic shear structures in *a* [201] extended mixed dislocations of SrTiO$_3$ unraveled by atomic-scale imaging using transmission electron microscopy and spectroscopy. *Faraday Discuss.* **213**, 245–258 (2019).

11. Hirth, J. P. & Lothe, J. *Theory of Dislocations*. (Krieger Pub Co, 1992).

12. Rodenbücher, C. *et al.* Mapping the conducting channels formed along extended defects in SrTiO$_3$ by means of scanning near-field optical microscopy. *Sci. Rep.* **10**, 17763 (2020).

13. Wrana, D., Rodenbücher, C., Bełza, W., Szot, K. & Krok, F. In situ study of redox processes on the surface of SrTiO$_3$ single crystals. *Appl. Surf. Sci.* **432**, 46–52 (2018).

14. Marrocchelli, D., Sun, L. & Yildiz, B. Dislocations in SrTiO$_3$: Easy To Reduce but Not so Fast for Oxygen Transport. *J. Am. Chem. Soc.* **137**, 4735–4748 (2015).

15. Szot, K. *et al.* Influence of Dislocations in Transition Metal Oxides on Selected Physical and Chemical Properties. *Crystals* **8**, 241 (2018).

16. Nakamura, A., Matsunaga, K., Tohma, J., Yamamoto, T. & Ikuhara, Y. Conducting nanowires in insulating ceramics. *Nat. Mater.* **2**, 453–456 (2003).

17. Szot, K., Speier, W., Carius, R., Zastrow, U. & Beyer, W. Localized Metallic Conductivity and Self-Healing during Thermal Reduction of SrTiO$_3$. *Phys. Rev. Lett.* **88**, 075508 (2002).

18. Okabe, T. H., Nakamura, M., Oishi, T. & Ono, K. Electrochemical deoxidation of titanium. *Metall. Mater. Trans. B* **24**, 449–455 (1993).

19. Bursill, L. A., Peng, J. & Fan, X. Structure and reactivity of atomic surfaces of barium titanate under electron irradiation. *Ferroelectrics* **97**, 71–84 (1989).

20. Szade, J., Psiuk, B., Pilch, M., Waser, R. & Szot, K. Self-neutralization via electroreduction in photoemission from $SrTiO_3$ single crystals. *Appl. Phys. A* **97**, 449–454 (2009).

21. Rodenbücher, C. *et al.* Electrically controlled transformation of memristive titanates into mesoporous titanium oxides via incongruent sublimation. *Sci. Rep.* **8**, 3774 (2018).

22. Wojtyniak, M. *et al.* Electro-degradation and resistive switching of Fe-doped $SrTiO_3$ single crystal. *J. Appl. Phys.* **113**, 083713 (2013).

23. Bussmann-Holder, A. *et al.* Unconventional Co-Existence of Insulating Nano-Regions and Conducting Filaments in Reduced $SrTiO_3$: Mode Softening, Local Piezoelectricity, and Metallicity. *Crystals* **10**, 437 (2020).

24. Koonce, C. S., Cohen, M. L., Schooley, J. F., Hosler, W. R. & Pfeiffer, E. R. Superconducting Transition Temperatures of Semiconducting $SrTiO_3$. *Phys. Rev.* **163**, 380–390 (1967).

25. Schooley, J. F., Hosler, W. R. & Cohen, M. L. Superconductivity in Semiconducting $SrTiO_3$. *Phys. Rev. Lett.* **12**, 474–475 (1964).

26. Binnig, G., Baratoff, A., Hoenig, H. E. & Bednorz, J. G. Two-Band Superconductivity in Nb-Doped $SrTiO_3$. *Phys. Rev. Lett.* **45**, 1352–1355 (1980).

27. Ueno, K. *et al.* Electric-field-induced superconductivity in an insulator. *Nat. Mater.* **7**, 855–858 (2008).

28. Reyren, N. *et al.* Superconducting interfaces between insulating oxides. *Science* **317**, 1196–1199 (2007).

29. Meevasana, W. *et al.* Creation and control of a two-dimensional electron liquid at the bare SrTiO$_3$ surface. *Nat. Mater.* **10**, 114–118 (2011).

30. Ge, J. F. *et al.* Superconductivity above 100 K in single-layer FeSe films on doped SrTiO$_3$. *Nat. Mater.* **14**, 285–289 (2015).

31. Rodenbücher, C. *et al.* Local surface conductivity of transition metal oxides mapped with true atomic resolution. *Nanoscale* **10**, 11498–11505 (2018).

32. Al-Zubi, A., Bihlmayer, G. & Blügel, S. Electronic Structure of Oxygen-Deficient SrTiO$_3$ and Sr$_2$TiO$_4$. *Crystals* **9**, 580 (2019).

33. Alexandrov, V. E., Kotomin, E. A., Maier, J. & Evarestov, R. A. First-principles study of bulk and surface oxygen vacancies in SrTiO$_3$ crystal. *Eur. Phys. J. B* **72**, 53–57 (2009).

34. Cuong, D. D. *et al.* Oxygen Vacancy Clustering and Electron Localization in Oxygen-Deficient SrTiO$_3$: LDA+U Study. *Phys. Rev. Lett.* **98**, 115503 (2007).

35. Li, W.-H. *et al.* Enhancement of superconductivity by the small size effect in In nanoparticles. *Phys. Rev. B* **72**, 214516 (2005).

36. Bezryadin, A., Lau, C. N. & Tinkham, M. Quantum suppression of superconductivity in ultrathin nanowires. *Nature* **404**, 971–974 (2000).

37. Ansermet, D. *et al.* Reentrant Phase Coherence in Superconducting Nanowire Composites. *ACS Nano* **10**, 515–523 (2016).

38. Migoni, R., Bilz, H. & Bäuerle, D. Origin of Raman Scattering and Ferroelectricity in Oxidic Perovskites. *Phys. Rev. Lett.* **37**, 1155–1158 (1976).

39. Bilz, H., Benedek, G. & Bussmann-Holder, A. Theory of ferroelectricity: The polarizability model. *Phys. Rev. B* **35**, 4840–4849 (1987).

40. Bussmann-Holder, A. The polarizability model for ferroelectricity in perovskite oxides. *J. Phys. Condens. Matter* **24**, 273202 (2012).

41. Roleder, K., Bussmann-Holder, A., Górny, M., Szot, K. & Glazer, A. M. Precursor dynamics to the structural instability in SrTiO$_3$. *Phase Transit.* **85**, 939–948 (2012).

42. Bussmann-Holder, A. A simple recipe to calculate the thermal conductivity of anharmonic crystals: the case of SrTiO$_3$. *Ferroelectrics* **553**, 26–35 (2019).

43. Bussmann-Holder, A. & Bishop, A. R. Intrinsic inhomogeneity as origin of incomplete ferroelectricity. *EPL Europhys. Lett.* **76**, 945 (2006).

44. Bussmann-Holder, A., Keller, H., Simon, A. & Bianconi, A. Multi-Band Superconductivity and the Steep Band/Flat Band Scenario. *Condens. Matter* **4**, 91 (2019).

45. Li, F. *et al.* Single-crystalline epitaxial TiO film: A metal and superconductor, similar to Ti metal. *Sci. Adv.* **7**, eabd4248 (2021).

**Acknowledgements**

We are thankful to C. Wood for proofreading the manuscript and to C. Korte for fruitful discussions. F.K. acknowledges the support of the Polish National Science Center (UMO-2018/29/B/ST5/01406).

**Author contributions**

K.S. prepared the samples and performed the LC-AFM measurements. K.Rog. conducted the experimental superconductivity analysis. A.B.-H. and H.K. performed the lattice dynamics analysis and G.B. the DFT simulations. K.S., K.Rog., G.B., W.S., C.R., K.Rol., F.K., H.K, and A.B-H. developed the model and wrote the manuscript with input from all of the coauthors.

**Competing interests**

The authors declare no competing interests.

**Supplementary information to**

**Filamentary superconductivity of resistively-switched strontium titanate**

K. Szot, C. Rodenbücher, K. Rogacki, G. Bihlmayer, W. Speier, K. Roleder, F. Krok, H. Keller, A. Simon and A. Bussmann-Holder

**Electro-reduction of SrTiO$_3$ crystals**

**Sample preparation**

Strontium titanate crystals (100) were purchased from CrysTec (Berlin, Germany). Samples with typical dimensions of 10 x 3 mm$^2$ and thicknesses of 0.5–1 mm were prepared by means of wire saw cutting. The samples were then cleaned in an ultrasonic bath in acetone, deionized water, and ethanol. The electrical connections were established using thin Pt wires in a 4-probe configuration[1] and, additionally, the contact sample/wires was pasted with conducting Pt suspension. Pre-heating in a vacuum at 200–300 °C for 0.5 h was employed for the removal of the organic solution of the Pt paste and water occluded on the surface of the sample.

**Electro-reduction apparatus**

Details of the aixDCA apparatus (aixACCT, Aachen, Germany), which was used for resistance measurements of the sample, are provided in Rodenbücher et al.[1]. The pressure in the vacuum chamber during the reduction process was $p < 10^{-7}$ mbar (the partial pressure of the oxygen was 2–3 orders of magnitude smaller than the base pressure). The sample was annealed to 350 °C and a voltage of 200 V, and a current compliance of 10 mA was applied. During the electro-reduction process, the potential drop between the inner electrodes was monitored electrometrically. In this manner, the total resistance of the sample (S), as well as the resistances of the bulk (B), anode region (A), and cathode region (C) were measured. For comparison, we also performed electro-reduction measurements in stepwise constant current mode with different currents of up to 10 mA leading to a higher initial voltage. For a polarization voltage higher than 200 V, the high ohmic input (Hi) of the electrometer (Keithley 6514, Keithley, Solon, USA) was connected to a voltage divider (1:100) consisting of high-impedance resistors ($10^{11}$ Ω/$10^9$ Ω +/- 1%). In all of the electro-reduction experiments, the current measurement was performed with a Keithley 6430 subfemtoampere source (Keithley, Solon, USA).

**Electro-reduction experiment**

The typical progression of the electro-reduction at a constant polarization voltage as a function of time is displayed in Figure S1a. The evolution of the resistance in stepwise constant current mode is shown in Figure S1b. In this mode, the current was increased in steps from 1 nA to 1 mA. As a criterion for switching to the next current step, a threshold voltage of 1 V was selected. Although the resistance decreased nearly continuously in constant voltage mode, the decrease in the resistance when using the stepwise constant current mode reached the maximum progression when the current was switched to 0.1, 1, 10 mA. It can be observed that in a relatively short time period, the resistance in all regions near the cathode, in the bulk, and near the anode was reduced by orders of magnitude. It should be noted that the time-dependent resistance changes during electro-reduction for different samples (e.g., from other suppliers or different batches) can exhibit fully dissimilar behavior, despite the same experimental parameters being present. This "undefined behavior" relates to the influence of different growth conditions and sample preparation (e.g., polishing methods) on the properties of the crystals and, in particular, on the electrical transport phenomena (for details, see Szot et al.[2]). In order to analyze the ex-corporation of oxygen during electro-reduction, we simultaneously measured the effusion of oxygen using a quadrupole mass spectrometer[2]. The mass spectrometric analysis revealed that during electro-reduction in stepwise constant current mode, an outflow of oxygen from the sample of the order of $10^{15}/cm^3$ was present. The oxygen outflow reached its maximum in the first stage of electro-reduction when the current was progressively increased. Nevertheless, after a few minutes, the oxygen concentration in the chamber decreased to the level it had been prior to the commencement of the electro-reduction. The extremely low oxygen concentration of only $10^{15}/cm^3$ that accompanied the electro-reduction process was calculated assuming that the oxygen vacancies were homogeneously generated in bulk. However, the electro-reduction of $SrTiO_3$ crystals was found to be fully inhomogeneous (see Figure 1 in the main manuscript) and restricted only to the dislocation cores arranged in a hierarchical network in the surface region. For such a selective de-oxidation generating a three-dimensional network of filaments, the actual local defect concentration in the reduced dislocation core was dramatically higher and can be assumed to be in the range of $10^{20-21}/cm^3$. Although electro-reduction can result in a decrease in the sample's original resistivity by 6–7 orders of magnitude (cf. Figure S1), this large drop in resistivity does not mean that the entire sample was converted into a metallic state. Therefore, the resistance dependence as a function of the temperature was analyzed after electro-reduction, as is shown in Figure S2. Two different electro-reduced crystals were compared. Figure S2a reveals a case, in which conversion to the metallic state upon electro-reduction in step-wise constant current mode ($I = 1$ nA - 10 mA, $T = 320$ °C, vacuum) only occurred in the bulk and the near-cathode region, as identified by the increasing resistance with temperature. Still, the near-anode region exhibited a behavior typical for a semiconductor. The investigation of the second sample, which was electro-reduced in constant voltage mode ($U = 300$ V, $I_{comp} = 10$ mA, $T = 300$ °C, vacuum) revealed metallicity in all regions (Figure S2b). This indicates that metallic filaments formed in the entire surface layer, guaranteeing that sufficient current can flow through the sample, even at low temperatures. Hence, this sample was selected for the superconductivity investigation.

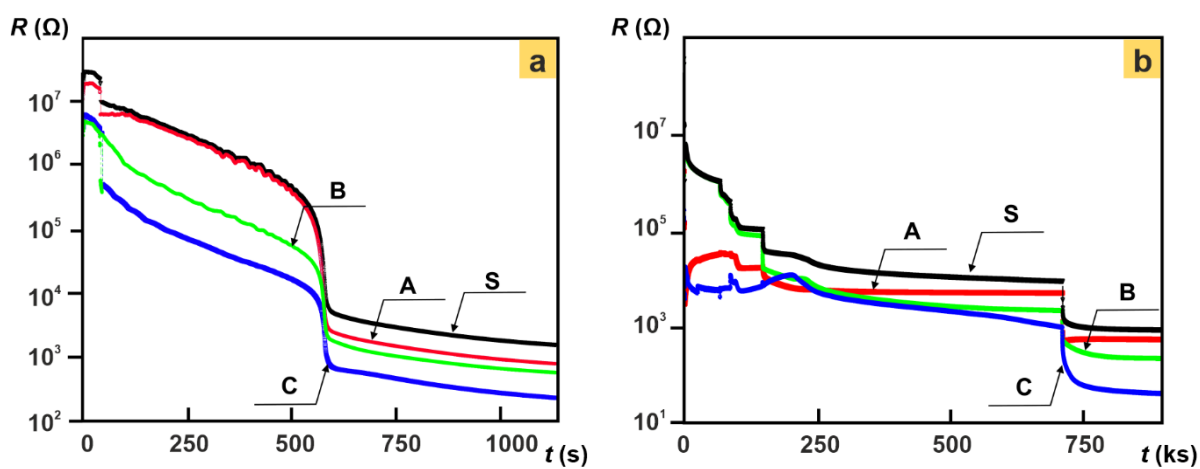

**Figure S1 Progression of the electro-reduction of SrTiO$_3$ (100) crystals ($T = 350$ °C, vacuum) for two polarization modes. a**) Constant voltage (200 V); **b**) stepwise constant current: 0↔11s:10$^{-9}$A, 11↔48s:10$^{-8}$A, 48↔160s:10$^{-7}$A, 160↔6928s:10$^{-6}$A, 6928↔8905s:10$^{-5}$A, 8905↔14820s:10$^{-4}$A, 14820↔71258s:10$^{-3}$A, and 71258↔end:10$^{-2}$A. The letters above the curves are abbreviations for: S-whole sample, B-bulk, A-region close to the anode, and C-region close to the cathode.

Despite the clear indication that the low-temperature (350 °C) electro-reduction under vacuum relates to the electrically-induced de-oxidation of the core of the filaments, it may be asked whether the thermal treatment by Joule heating under vacuum conditions can also contribute to the decrease in macroscopic resistance. Through measurement, we determined that during electro-reduction with 10 mA, the increase in the sample's resistance due to Joule heating was well below 100 °C[3]. Therefore, we checked the influence of thermal conditions on the potential self-doping of our crystals. A pristine crystal was heated in air to 380 °C. Then, the chamber was evacuated and the temperature was maintained for four hours before the sample was cooled to room temperature. Figure S3 shows that the resistance measured versus the temperatures exhibited nearly identical semiconducting behavior during heating in air and cooling in a vacuum proving that no significant thermal reduction occurred at this temperature.

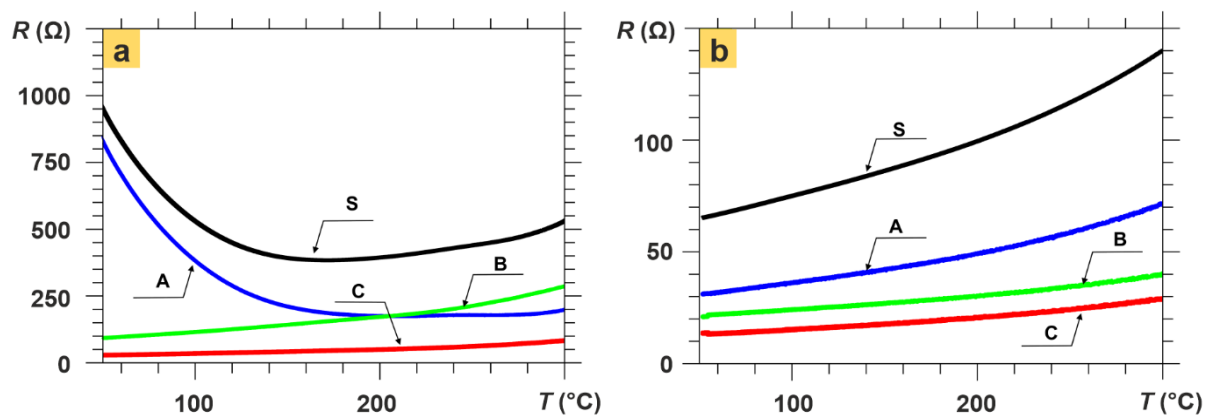

**Figure S2 The temperature-dependence of the electrical resistance for different regions of the electro-reduced crystal. a**) Metallic transformation of the region close to the anode only and **b**) full transformation of the entire sample into a metallic state (the notation S, B, A, C is the same as in Fig. 1).

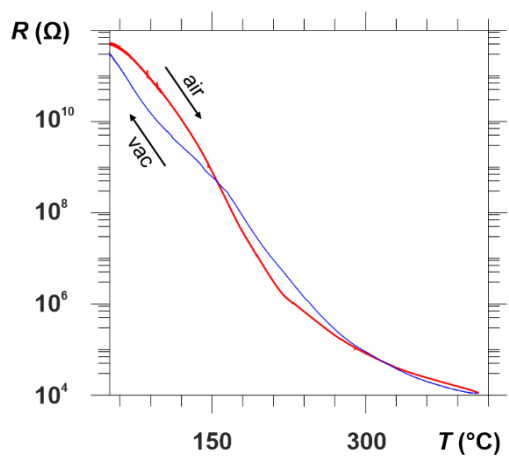

**Figure S3 Temperature-dependence of the resistance of SrTiO$_3$ crystals during heating in air and subsequent cooling under vacuum conditions.** At the maximal temperature of approximately 400 °C, the sample was annealed for 4 h under vacuum conditions.

### LC-AFM investigation of the electrical conductivity on the nanoscale

Following electro-reduction, the samples were cooled to room temperature under vacuum conditions, but their transfer from the preparation chamber to the cryostat was undertaken ex-situ. Although this transfer was performed at room temperature, the freshly-prepared sample was exposed to the ambient atmosphere. To ensure comparability, the LC-AFM analysis was also performed after exposing the sample to ambient conditions (Figure 2 in the main text). The measurements themselves were performed under vacuum conditions at $10^{-5}$ mbar using a JSPM 4210 setup (JEOL, Tokyo, Japan). For the LC-AFM mapping of the electrical conductivity on the crystals' cross-sections (Figure 1 in the main text), the samples were cleaved in situ. In order to prevent post-oxidation of the freshly-cleaved surfaces, the cleavage process, and finally, the measurement, was carried out under vacuum conditions with the partial pressure of the oxygen reduced to $10^{-15}$–$10^{-17}$ mbar by the addition of $H_2$ to the chamber at a total pressure of $10^{-6}$ mbar. For the LC-AFM investigation, conducting Pt/Ir-coated tips (PPP-ContPt, Nanosensors, Neuchâtel, Switzerland) were used and biased with a few millivolts.

The LC-AFM maps were analyzed using the WinSPM software (JEOL, Tokyo, Japan). In order to reduce the noise level of the atomically-resolved conductivity maps, only a software low-pass filter was used (without FFT transformation), noting that on all of the LC-AFM maps with atomic resolution, the distribution of atoms was visible by the naked eye. In order to ensure that true atomic resolution was acheived, the LC-AFM data were only examined when the position of atoms (in the same region) was the same for several scans.

**Self-doping of SrTiO$_3$ crystals caused by thermal reduction**

Thus far, superconductivity in stoichiometric SrTiO$_3$ has only been explored for thermally-reduced crystals. Hence, for the analysis of the superconducting properties of electrodegraded crystals, we were obliged, *per se*, to compare the properties of electrodegraded crystals with thermally-prepared ones. To avoid an influence of the sample preparation process due, e.g., to different growth parameters, polishing methods and concentrations of impurities on the progression of both the reduction and electrodegradation processes, the experiments were performed on samples prepared from the same crystal.

For the thermally-induced self-doping of the crystal, we chose an annealing temperature of about 800 °C and a pressure of $10^{-7}$ mbar. In contrast to an *uncontrolled* reduction process, as would be caused by placing the crystals in the hot area of the furnace in the vacuum chamber and waiting until equilibrium was reached, we analyzed the progression of thermal reduction by monitoring the resistance as a function of the reduction time (Figure S4-left) as described by Rodenbücher et al.[1]. It can be seen that the resistance initially decreases and then increases again as a result of self-healing[4]. The presence of the minimum on the curve $R(t)_{T\text{const},p\text{const}}$ (marked with an arrow) clearly shows that it is necessary to stop the reduction after a certain time in order to obtain the maximum reduction or maximum concentration of self-doped carriers. It is surprising that this thermal reduction, which transformed the sample into a metallic state (see Figure S4, right) and finally into a superconducting crystal (Figure 3 in the main text), only results in $10^{14}$/cm$^3$ oxygen atoms leaving the entire sample[2,4].

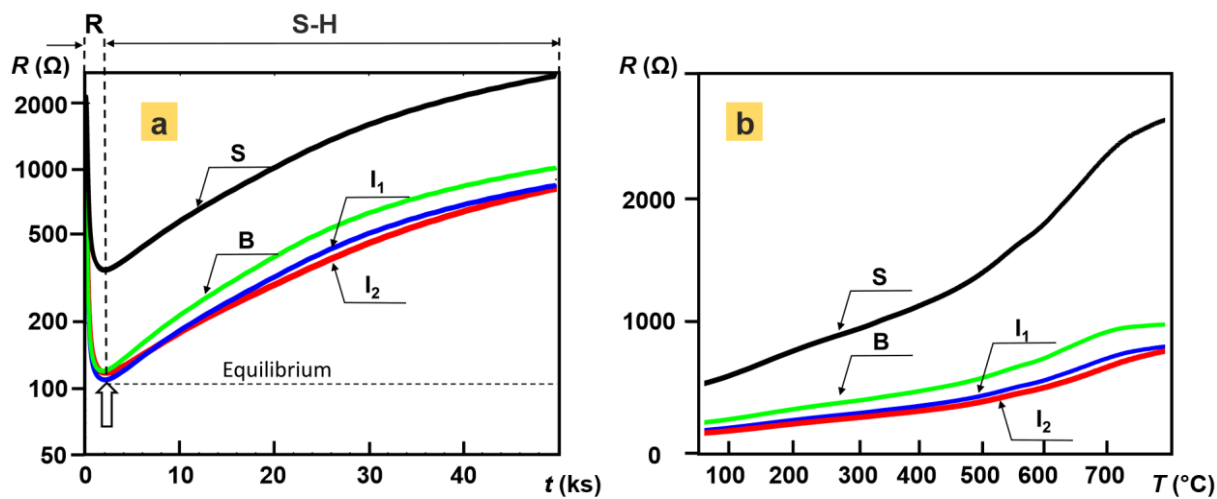

**Figure S4 Progression of the thermal reduction of an SrTiO$_3$ crystal at 800 °C in a vacuum for different regions of the samples. a)** Resistance as a function of reduction time (S-whole sample, B-bulk, and $I_1$-$I_2$-Interfaces close to electrodes). The position of the white arrow marks the optimal reduction state (R) induced in the crystal. The prolonged reduction does not allow for the reaching of the equilibrium of the defects, but leads to the self-healing process (S-H); in the depicted process, the resistance increases by a factor of > 5 (after a long reduction time) relative to the maximum reduction state reached after only a few minutes[4]. **b)** thermal dependence of the resistivity of a thermally reduced crystal revealing that the reduction process results in the conversion of all regions in the crystals into the metallic state.

Hence, the macroscopically-determined increase in resistivity cannot be interpreted as the effect of self-doping of the entire crystal, but is an effect of the selective removal of oxygen from the dislocation network in the surface region[2,5,6]. In this regard, thermal reduction and electrodegradation are similar, as both processes occur, in principle, in the hierarchical network in dislocations. Of course, based on the point defect chemistry for the reduced ternary oxides at low oxygen activity (here, the vacuum) levels, it can be assumed that "classical doping" of the matrix should occur. It is assumed that the two-fold positive charge of oxygen vacancies is compensated by electrons under reducing conditions, thus turning the matrix into an n-type semiconductor. As the LCAFM measurement of the electrical conductivity of the matrix (e.g., between filaments and in the bulk[7,8]) in thermally-reduced crystals exhibits no decrease in resistivity, we can conclude that such a potential homogeneous doping with electrons is not relevant for the nano- and macroscopic transition to the metallic state. The matrix of thermally-reduced crystals is just as isolating as in the case of stoichiometric crystals[9]. In-situ 4-tip STM investigations[6] show that the thermal reduction that occurs in the surface layer only provides a further proof of the inhomogeneity of the thermal reduction. The same conclusion regarding the dominant role of the surface layer can be reached by analyzing data from so-called resistivity tomography (see Figure S5). Using the micro-Valdes four-point technique, we analyzed the contribution of the different regions to the total resistance of the optimally-reduced crystal after the subsequent removal of the layer with a thickness of a few µm. It can be clearly seen that resistance in the top layer is the lowest, which indicates that the last 10–20 µm of the reduced crystals channel the current flow. We reached a similar conclusion after analyzing from the LC-AFM maps the integrated contribution of filaments in the surface region of 10–20 µm in thickness in thermally-reduced (see [2]) and electro-reduced crystals (see Figure 1c in the main manuscript).

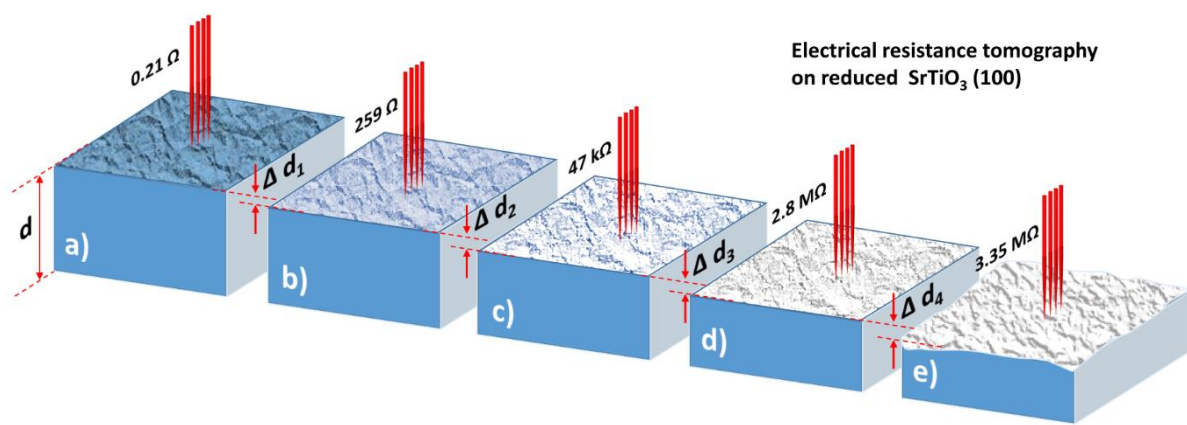

**Figure S5 Resistance tomography obtained with the micro-Valdes method for thermally-reduced SrTiO₃ crystals. a**) Resistance of the reduced crystal with the original surface (d = 0.5 mm). After the successive removal of 10 µm-thick layers, the resistance decreased by three orders of magnitude for **b**), and five orders of magnitude for **c**). After further grinding of the sample by 20 µm (each **d**) and **e**)), the resistance of the rest of the reduced crystal (representing the bulk resistance) was seven orders of magnitude higher than for the crystals in the upper part of the surface region. Note that the crystals for the topographic measurement were optimally-reduced (i.e., the reduction state reached its maximum in the crystals).

**Superconducting properties measurements**

Electrical resistance measurements were performed for rectangular samples with dimensions of 1.0 x 3.1 x 10 mm$^3$, using the physical properties measurement system (PPMS Model 7100, Quantum Design, San Diego, USA) equipped with a 14 T superconducting magnet. The four-point technique was used to measure the long-term resistance, *R*, with a suitably high level of accuracy. Current leads (silver wires with a diameter of 0.1 mm) were glued with silver epoxy to the end faces of the samples in order to obtain a homogenous current distribution in the central area where voltage leads; silver wires with a diameter of 0.05 mm were glued at a distance of $l \approx 5$ mm, which varied slightly for individual samples. The obtained contacts had a resistance below 1 Ω for current leads and 20–50 Ω for voltage leads (~ 0.5 mm-wide silver-epoxy strips) and were stable over time. For the resistance measurements, the electric transport option (ETO) was used with an AC current of 18–21 Hz frequency and amplitude of 0.05–0.2 mA. For these conditions, the resistance in the normal state was frequency- and amplitude-independent. The resistance in the magnetic field was measured at increasing temperatures for a fixed field (magnet persistent mode) or, at increasing and decreasing fields, for fixed temperatures. The temperature was changed at a rate of 0.02 K/min, for which no significant *R*(*T*) hysteresis was observed for the up and down sweeps.

**DFT simulation**

We performed density functional theory calculations in the generalized gradient approximation[10], employing the full potential linearized augmented planewave method[11]. In order to correct the band gap of SrTiO$_3$, we used a Hubbard *U* correction, as proposed in[12]. For the simulation of the extended defect, three unit cells of SrTiO$_3$ were removed from a 6 x 5 x 1 supercell with an additional oxygen vacancy in the lower left of the defect (see the inset of Figure 3a in the main text). The defect-induced states were energetically located at the bottom of the conduction band to estimate their spatial distribution and the charge density was integrated into the atomic sphere of 1.9 Å around the Ti sites (the red bars in Figure 3b). The local polarizations in each unit cell (the yellow arrows in Figure 3a and black bars in Figure 3b) were calculated from formal charges of +2 and –2 for Sr and O, respectively, and their relaxed positions in the Ti-centered unit cell.

**Supplementary References**

1. Rodenbücher, C., Korte, C., Schmitz-Kempen, T., Bette, S. & Szot, K. A physical method for investigating defect chemistry in solid metal oxides. *APL Mater.* **9**, 011106 (2021).

2. Szot, K. *et al.* Influence of Dislocations in Transition Metal Oxides on Selected Physical and Chemical Properties. *Crystals* **8**, 241 (2018).

3. Rodenbücher, C. *et al.* Current channeling along extended defects during electroreduction of $SrTiO_3$. *Sci. Rep.* **9**, 2502 (2019).

4. Szot, K., Speier, W., Carius, R., Zastrow, U. & Beyer, W. Localized Metallic Conductivity and Self-Healing during Thermal Reduction of $SrTiO_3$. *Phys. Rev. Lett.* **88**, 075508 (2002).

5. Rodenbücher, C., Wojtyniak, M. & Szot, K. Conductive AFM for Nanoscale Analysis of High-k Dielectric Metal Oxides. in *Electrical Atomic Force Microscopy for Nanoelectronics* (ed. Celano, U.) 29–70 (Springer International Publishing, 2019). doi:10.1007/978-3-030-15612-1_2.

6. Leis, A. *et al.* In-situ four-tip STM investigation of the transition from 2D to 3D charge transport in $SrTiO_3$. *Sci. Rep.* **9**, 2476 (2019).

7. Szot, K., Speier, W., Bihlmayer, G. & Waser, R. Switching the electrical resistance of individual dislocations in single-crystalline $SrTiO_3$. *Nat. Mater.* **5**, 312–320 (2006).

8. Waser, R., Dittmann, R., Staikov, G. & Szot, K. Redox-Based Resistive Switching Memories - Nanoionic Mechanisms, Prospects, and Challenges. *Adv. Mater.* **21**, 2632–2663 (2009).

9. Rodenbücher, C. *et al.* The Electronic Properties of Extended Defects in $SrTiO_3$—A Case Study of a Real Bicrystal Boundary. *Crystals* **10**, 665 (2020).

10. Perdew, J. P., Burke, K. & Ernzerhof, M. Generalized Gradient Approximation Made Simple. *Phys. Rev. Lett.* **77**, 3865–3868 (1996).

11. Wimmer, E., Krakauer, H., Weinert, M. & Freeman, A. J. Full-potential self-consistent linearized-augmented-plane-wave method for calculating the electronic structure of molecules and surfaces: $O_2$ molecule. *Phys. Rev. B* **24**, 864–875 (1981).

12. Al-Zubi, A., Bihlmayer, G. & Blügel, S. Electronic Structure of Oxygen-Deficient $SrTiO_3$ and $Sr_2TiO_4$. *Crystals* **9**, 580 (2019).